# "Unusual Rainbow and White Rainbow"

## A new auroral candidate in oriental historical sources

**Affiliations**

Hisashi Hayakawa (1), Hiroaki Isobe (2, 3), Akito Davis Kawamura (4), Harufumi Tamazawa (4), Hiroko Miyahara (5), Ryuho Kataoka (6, 7)

(1) Graduate School of Letters, Kyoto University, Sakyo, Kyoto, 606-8501, Japan

(2) Integrated Advanced Institute for Human Sustainability, Kyoto University, Sakyo, Kyoto, 606-8306, Japan

(3) Unit of Synergetic Studies for Space, Kyoto University, Sakyo, Kyoto, 606-8502, Japan

(4) Kwasan Observatory, Kyoto University, Yamashina, Kyoto, 607-8471, Japan

(5) Musashino Art University, kodaira, Tokyo, 187-8505, Japan

(6) National Institute of Polar Research, Tachikawa, Tokyo, 190-8518, Japan

(7) Department of Polar Science, SOKENDAI, Japan

**Abstract**

Solar activity has been recorded as auroras or sunspots in various historical sources. These records are of much importance for investigating both long-term solar activities and extremely intense solar flares. In previous studies, they were recorded as "vapor," "cloud," or "light," especially in oriental historical sources; however, the terminology was not discussed adequately and is still quite vague. In this paper, we suggest the possibility of "unusual rainbow" and "white rainbow" as candidates of historical auroras in oriental historical sources and examine if it is probable. This discovery will help us to make more comprehensive historical auroral catalogues and require us to add these terms to auroral candidates in oriental historical sources.

**1. Introduction**

The magnetic field of the sun causes various spectacular phenomena in the solar atmosphere and the interplanetary space. Evidence of solar magnetic activities in the pretelescopic era can be found in the records of aurora and sunspot observations in various historical sources (Eddy 1980; Nakazawa et al. 2004; Usoskin 2013; Hayakawa et al. 2015; Neuhäuser & Neuhäuser 2015). These types of records are of great importance to investigate the long-term variability of solar activities as well as the probability of occurrence of extremely intense solar flares (Owens 2013; Araki 2014). Recently, solar flares are attracting increasing interests because of the discovery of several "super flares," i.e., flares whose total energy exceeds that of the known largest solar flares by orders of magnitudes, in solar-like G-stars (Maehara et al. 2012, Shibayama et al. 2013, Shibata et al. 2013) as well as the possible, but not yet confirmed, evidence of the occurrence of extremely intense solar flares in the past found in radioactive isotopes (Miyake et al. 2015).

One of the difficulties encountered in previous studies on past solar activities using historical literature was that ancient observers may have used different terminologies to record heavenly events about which they did not have scientific understanding. Kanda (1933) claimed that the term "red vapor (赤氣)" refers to an aurora, and compiled the first historical auroral catalogue by picking up several words with "vapor (氣)." Keimatsu (1970–76) conducted a more extensive survey of heavenly events including "all the luminous phenomena in the sky." As Keimatsu's catalogues seemed to include many "meteors," Yau (1995) extracted the list of "red vapors" in an attempt to remove such terms.

Because of the recent advances in digital databases of historical literature, it is becoming possible to search selected keywords more easily and completely than before. In our previous study (Hayakawa

2015), we used Scripta Sinica, a comprehensive database of Chinese official histories, to search for potential records of sunspots and auroras by using the keywords "vapor (氣)," "cloud (雲)," or "light (光)." in *Sòngshǐ* (宋史), a Chinese official history spanning from 961 to 1279.

Through such an investigation, we found an unknown word "unusual rainbow (虹蜺)" in the chapter of "red vapors" in the Astronomical Treatise of *Jiùtángshū* (舊唐書; hereafter noted as JT), a Chinese official history in the reign of *Táng* (唐) dynasty (618 to 907). As this "unusual rainbow" was included in the chapter of "red vapor," which is a well-known historical auroral term, it is reasonable to consider this term as a candidate for auroral record.

Moreover, we found that there are records of "white rainbow（白虹）" in the sky that are potentially aurora records. Previously, "white rainbow" was not considered as an aurora because its typical usage is "white rainbow pierced the sun (白虹貫日)," which strongly indicates atmospheric optics phenomena such as solar halos. However, as we will show later, there are a few records of "white rainbows" whose descriptions seem to be more consistent with auroras rather than with solar halos (figure 1). Historical pictures for auroras described as a bow above the sky such as one recorded in Ramus (1745) also support our hypothesis (figure 2). Chapman et al. (2015) also mentioned a possibility that white rainbow in 757 can be an aurora.

In this paper, we collect contemporary "unusual rainbows" and "white rainbows" to re-examine if they can be considered as new auroral candidates. We note that the purpose of this study is to collect not only well-known SAR-arc type or broad-band electron type mid-latitude/low-latitude auroras, which were already investigated in the space age (c.f. Shiokawa et al., 2004), but also to learn what actually happens when unprecedented magnetic storms are observed in the space age, including "whitish" discrete-type auroras overhead, which might be witnessed even in mid-latitude/low-latitude areas and recorded in old literature.

**2. Method**

We examined Chinese official histories in the reign of *Táng*. The contemporary Chinese official chronicles were compiled twice: first as *Jiùtángshū* (舊唐書; JT) and later as *Xīntángshū* (新唐書; XT). The former was compiled in 945 and the latter in 1060, and both cover the same period (618-907). We surveyed the passages that included "white rainbow (白虹)" or "unusual rainbow (虹蜺)" from JT and XT by using the mechanical retrieval system of the digital database Scripta Sinica (新漢籍全文) (http://hanchi.ihp.sinica.edu.tw/ihp/hanji.htm), and examined them philologically and scientifically to check whether they refer to aurora observations.

## 3. Results

XT contains several treatises, including *Tiānwénzhì* (天文志), the treatise of astronomical events, and *Wǔxíngzhì* (五行志), the treatise of natural events and their interpretation based on traditional Chinese *Wǔxíng* philosophy. In XT, "white rainbows" appear in the sections of "solar omens" (日變)[1] and "lunar omens" (月變)[2] in the treatise of *Tiānwénzhì* and the section of "unusual rainbows" (虹蜺) in the treatise of *Wǔxíngzhì*.

The sections of "solar omens" and "lunar omens" describe the records of unusual events associated with the sun and the moon, respectively. Most "white rainbows" in these sections are described as "white rainbow pierced the sun" (白虹貫日) or "rainbow pierced the moon" (有虹貫月). It is likely that the "white rainbows" described in these sections are solar/lunar halos or similar atmospheric optics phenomena, as interpreted in normal dictionaries. We exclude such phenomena from our investigation.

On the contrary, the usage of "white rainbow" in the section of "unusual rainbow" in the treatise of *Wǔxíngzhì* seem to include the description of phenomena that are not related to solar/lunar halos or usual rainbows. We found nine records of "white rainbow" and one record of "unusual rainbow" in the section of "unusual rainbow" in *Wǔxíngzhì*.

---

[1] We found 1 in JT and 17 in XT with dates.
[2] We found 1 in JT with a date.

In JT, there are no independent sections of "solar omens," "lunar omens," or "unusual rainbows." Therefore, we searched for all records that included "rainbow," and then manually removed those that were explicitly associated with the sun or the moon. Two records of "white rainbow" and one record of "unusual rainbow" were found. In total, there were eleven records of "white rainbow" and two records of "unusual rainbow" in JT and XT that are potential candidates of auroral observations. They are listed below. The format of the example is described as follows: their ID number, their date of observation, reference of their original texts, their original texts in classical Chinese, and their translations into English.

**White and unusual rainbows in JT (舊唐書)**

JT#1//868.7.29（JT, Yìzōng, p663）咸通九年…七月戊戌，白虹橫亙西方。

On 29th July 868, a white rainbow lied across westward.

JT#2//882.10.xx（JT, Xīzōng, p725）中和二年十月…壬辰夜，白虹見西方。

On xx[3] October 882, at night, a white rainbow was observed westward.

JT#3//710.7.9（JT, Astronomy II, Red vapor, p1324）唐隆元年六月八日，虹蜺竟天。

On 9th July 710, an unusual rainbow filled up the heaven.

**White and unusual rainbows in XT (新唐書)**

XT#1//618~（XT, Five Elements III, 950）武德初，隋將堯君素守蒲州，有白虹下城中。

In the early *Wǔdé* period (CE 618 to 626), *Yáo Jūnsù*, a general of *Suí*, guarded *Púzhōu,* and then a white rainbow descended into the city.

XT#2//710.7.9（XT, Five Elements III, 950）唐隆元年六月戊子，虹蜺亙天。蜺者，斗之精。占曰：「后妃陰脅王者。」又曰：「五色迭至，照于宮殿，有兵。」

On 9th July 710, an unusual rainbow filled up the sky. A rainbow is a spirit of the Plough. Fortune-tellers said, "The Empress secretly scare the ruler." They also said, "When five colors arrive one by one and illuminate a palace, wars break out."

XT#3//712.7（XT, Five Elements III, 950）延和元年六月，幽州都督孫佺帥兵襲奚，將入賊境，有白虹垂頭于軍門。占曰：「其下流血。」

---

[3] This "xx" represents an error date in records.

In July 712, *Sūn Quán*, a governor-general of *Yōuzhōu*, led his army and attacked *Xī*. When he was about to enter the hostile territory, a white rainbow descended to the gate of their camp. Fortunetellers said, "Below, it bleeds."

XT#4//757.2.20（XT, Five Elements III, 950）至德二載正月丙子，南陽夜有白虹四，上亙百餘丈。

On 20th Feb. 757, four white rainbows were observed at night in *Nányáng* and they extended to a length of more than hundred *zhàng*.

XT#5//819.1.6（XT, Five Elements III, 950）元和十三年十二月丙辰，有白虹闊五尺，東西亙天。

On 6th January 819, a white rainbow, as wide as 5 *chǐ*, filled up the heaven and lied across east to west.

XT#6//844.2.17（XT, Five Elements III, 950）會昌四年正月己酉，西方有白虹。

On 17th February 844, a white rainbow was observed westward.

XT#7//861.6.xx（XT, Five Elements III, 950）咸通元年七月己酉朔，白虹橫亙西方。

On xx June 861, a white rainbow was observed lying across westward.

XT#8//868.7.29（XT, Five Elements III, 950）咸通…九年七月戊戌，白虹橫亙西方。

On 29th July 868, a white rainbow was observed lying across westward

XT#9//886.10（XT, Five Elements III, 950）光啓二年九月，白虹見西方。

In October 886, a white rainbow was observed westward.

XT#10//886.11.xx（XT, Five Elements III, 950）光啓二年…十月壬辰夜，又如之（白虹見西方）。

On xx November 886, it was also like this (a white rainbow was seen westward).

The information on observation time is of crucial importance because auroras appear at night and usual rainbows appear during daytime. Unfortunately, most records lack the information on observation time. In the lists mentioned above, only JT#2, XT#4, and XT#10 describe that the "rainbows" were observed at night.

Moreover, directions and motions in the sky are of importance to examine the likelihood of the occurrence of auroras. In JT#3, XT#2, and XT#5, the rainbows are described as "filled up the sky" (竟天 or 亙天)[4]. In particular, XT#5 is described as "filled up the heaven and lied across east to west" (東西亙天), indicating that the "white rainbow" spread across a large area of the sky. JT#1, JT#2, XT#6, XT#7, XT#8, XT#9, and XT#10 are described to appear in the western sky. There are two other descriptions regarding the motion of "white rainbow": "white rainbow descended into the city." in XT#1, and "white rainbow descended to the gate of their camp" in XT#3, for which we have no interpretations that can be associated with auroras.

---

[4] "亙天" has similar meaning with "竟天".

Finally, there are two records describing the size of "white rainbow": "they extended to a length of more than hundred *zhàng*" in XT#4 and "as wide as 5 *chǐ*" in XT#5. We do not know how to convert them to a suitable size in the plane of the sky.

**4. Discussion**

**4.1 Are they auroras?**

In this section, we argue that the historical records of "white rainbow" and "unusual rainbow" include the observations of auroras. It has been generally believed that "white rainbows" indicate atmospheric optics events associated with the sun or the moon, as their common usages are "white rainbow pierced the sun" (白虹貫日) or "rainbow pierced the moon" (有虹貫月). As mentioned above, however, XT had independent sections for solar omens and lunar omens, in which "white rainbow pierced the sun" or "rainbow pierced the moon" were included, respectively. Therefore, it is reasonable to consider that the "white rainbows" and "unusual rainbows" described in the section of "unusual rainbows" are different from the atmospheric optics events of solar or lunar origin.

Auroras are faint, and hence, they must have been observed at night. Although most records lack the information on time, there are three "white rainbows" that were explicitly recorded as being observed at night. It is unlikely that these records are referring to usual rainbows or the atmospheric optics of the sun. There are no explicit records of daytime observations of "white rainbows" or "unusual rainbows" in our list.

In the case they are observed in the daytime, we expect their nature to be "fogbow". Fogbows appear when the sunlight is reflected by drops of fog, which have a diameter smaller than 0.05 mm (Vaquero et al. 2002). They have no color and look whitish, and thus are likely to be called "white rainbow." There are some reports inliterature including Carrington (1871) and Vaquero et al. (2002)

Moreover, the observations have limited information on the probability of the nature of records. The phenomena of atmospheric optics of the moon are usually observed during fall-winter and much less during the summer. However, another possible non-auroral candidate of "white/unusual rainbow" besides atmospheric optics, i.e., noctilucent clouds, preferentially occur during the summer.

The size of the rainbows in the sky puts a stronger constraint. As mentioned in the previous section, there are several records of "white/unusual rainbows" that "filled up the sky." Noctilucent clouds, paraselene, or usual rainbows cannot spread across a large area of the sky. Noctilucent clouds appear only during twilight. Theoretically, the structures of paraselene and parhelic circles have the widest distribution; however, they appear horizontally at the same altitude as the moon instead of spreading across the sky through the region around the zenith. Therefore, the recorded phenomena described as "filling up the sky" are likely to imply aurora. In particular, "a white rainbow as wide as 5 *chǐ* filled up the heaven and lied across east to west" (XT#5) gives the impression that a hazy light covered the entire sky, invoking auroras. One may think that hazy clouds in the night can also "fill up the sky." Although we cannot exclude that possibility, we think that the phenomenon of a cloudy sky is so common and obvious that it would not be recorded in the section of "unusual rainbows."

Regarding the color, the auroras are not really "white." They exhibit a green color from the forbidden line of atomic oxygen at 537.7 nm at mid altitudes (100–150 km), and red color from the forbidden line of atomic oxygen at 630.0 nm at high altitudes (>200 km). In very intense auroras, blue, violet, or pink colors are observed at low altitudes (80–100 km) that originate from the molecular nitrogen lines in the blue and red parts of the spectrum. However, human eyes can correctly recognize a color when a target is bright enough. When observing a very faint aurora, i.e., an aurora whose flux of energetic electron precipitations is not large, the color can be recognized as whitish because the target is not bright enough. In addition, the color of an aurora actually changes vertically because of atmospheric scattering. For example, when a green aurora appears near the horizon, it may appear yellow. Moreover, when recording colors, ancient Chinese observers used five colors, i.e., white, red, blue, yellow, and black, which corresponds to the traditional concept of *Wǔxíng* (五行) or Five Elements, namely metal, fire, wood, soil, and water, respectively. Therefore, it is reasonable to assume that green or faint yellow auroras were expressed as white as they do not have "green" as a base color in their traditional concept.

It should be noted that low-latitude auroras are usually observed as red auroras in the direction of higher latitudes (north of the northern hemisphere). Observations of white/green/yellow auroras and auroras that "fill up the sky" strongly indicate that the auroral oval extended to a low latitude of 35N, where the Chinese capitals in the *Táng* era were located. Such low-latitude auroras are rare, but they did occur, e.g., during very intense geomagnetic storms such as the one that occurred in 1989 (Allen et al. 1989). Hayakawa et al. (2015) indicated that "white" auroras in the past might have been observed more easily compared to the present because the geomagnetic latitude of China in the *Sòng* era could be higher than that in the present (Butler 1992).

**4.2 Simultaneous observations**

Since auroras are a global phenomenon, the simultaneous observations of aurora-like phenomena at distant locations strongly support their interpretation as auroras. Examples of such simultaneous observations in the historical literature were reported by Willis & Stephenson (1999). These observations include "a red and white cloudy vapor" in China and "red vapors" in Japan, on 8th March 1582, and "flames" in China and "red and white vapor" in Japan on 2nd March 1653.

Thus far, we have not found any simultaneous observations of the "white/unusual rainbows" in the *Táng* dynasty listed above. However, we have found records of "white vapor" observations during the *Qīng* dynasty; corresponding observations of aurora-like phenomena have been observed in Europe, as catalogued by Fritz (1983). For example, on 17th September 1770, "white vapor" was observed at *Féichéng* (肥城) in China, which was simultaneously observed as "red vapors" in various cities in Japan (Kanda 1933) and auroral candidates in 7 places in western Europe (Fritz 1873). This example can be regarded as evidence that "white" is associated with a low-latitude aurora.

Another interesting case is the simultaneous observation on 30th July 1363, in China and in Japan (Willis & Stephenson 1999). Chinese astronomers in *Yuán* (元) dynasty reported an observation at *Jiàngzhōu* (絳州)[5] and stated as (a).

---

[5] *Jiàngzhōu*: present 山西省運城市新絳縣 (N 35°36′, E 111°13′).

(a) 至正二十三年…六月丁巳，絳州日暮有紅光見于北方，如火，中有黑氣相雜，又有白虹二，直衝北斗，逾時方散。（*Yuánshǐ*, Five Elements II, p1103）

On 30th July 1363, at sunset there was a red light in the north, like a fire, containing black vapor and two white rainbows, which directly hit the Plough and disappeared in some time.

Note that the "white rainbows" are associated with the "red light in the north like a fire," which strongly indicates an aurora. The interpretation as an aurora is supported by the record of a similar observation in Japan on the same day. As reported by Kanda (1933), *Sanjo Kimitada* (三条公忠), a contemporary minister of interiors in the medieval Japanese imperial court at Kyoto, left a diary with entries spanning from 1361 to 1383 and reported as follows.

(b)正平十八年…六月十九日，入夜艮幷北方如遠所燒亡，火光不知何故，或説，炎旱之瑞云々。（*Gogumaiki*, I, p62）

On 30 July 1363, at the beginning of night, it was like an inferno from north-east to north. The reason for this flame is not known and it was also said that this is an omen of a burning drought.

This case provides another support for the use of a "white rainbow" to express an aurora observation.

**4.3 Scientific and academic impacts**

Combining the above arguments, we believe that it is reasonable to consider "white rainbows" and "unusual rainbows" as candidates of auroral observations in future surveys of historical aurora records. Of course, one should note that these candidate words may have been used to express different phenomena, such as atmospheric optics and clouds. To ascertain whether the aforementioned records are actually records of auroras, it is necessary to read the passages carefully; however, there would still remain inevitable uncertainties in the nature of the observed phenomena as well as in the associated information such as date, time, and colors.

Here we present two cases in which the records of "white/unusual rainbow" have potential scientific/academic impacts. The first case is from a historical point of view. According to Kanda (1933), a "red vapor" on 30 December 620 was the first recorded aurora in Japan. However, we found a record of "rainbow at night" observed on June 459:

(c)　（雄略天皇三年四月…）天皇疑皇女不在，恒使闇夜東西求覓。乃於河上虹見如蛇、四五丈者。掘虹起処、而獲神鏡。（*Nihonshoki*, I, p467）

(In June 459…) The Emperor suspected the princess to be not there and made his servants search for her everywhere from east to west on a dark night. Above the river, a rainbow like a serpent as long as 4 to 5 *shaku* was found. Searching for the rainbow, he found a divine mirror.

If this is indeed a record of an aurora, it becomes the oldest record in Japan.

The second, more scientific case is the occurrence of extremely intense solar flares in the past. Recently, Miyake et al. (2012) found a rapid increase in the cosmogenic carbon-14 content in a tree ring in 775, which indicates the presence of very intense cosmic ray flux into the Earth's atmosphere. Mekhaldi et al. (2015) examined beryllium-10 and chlorine-36 from both Arctic and Antarctic ice cores and supported Miyake's discovery. Its origin is still controversial (a nearby supernova (Miyake et al. 2012), a gamma ray burst (Hambaryan and Neuhäuser, 2013; Pavlov et al. 2013), and a cometary impact on Earth (Liu et al. 2014)), but one possibility is extremely intense "superflare" on the sun (Shibata et al. 2013). Historical records of a low-latitude aurora in 775 can be supporting evidence for the solar origin of the carbon-14 event; these records are however not conclusive.

Although previous studies for Japanese chronicles found no such record in contemporary Japanese

official history *Shokunihongi*, we found a new auroral candidate recorded as "white rainbow":

(d) 丙午。白虹竟天。（*Shokunihongi*, p421）

On 16 July 775, a white rainbow filled up the heaven.

This one is philological and quite similar to "white rainbows" in XT, and hence, it may be an auroral record.

Carbon-14 is absorbed in trees by photosynthesis in the form of carbon dioxide, after circulating in the carbon cycle. Based on the measurement of carbon-14 in tree rings from both northern and southern hemispheres, Guettler et al. (2015) estimated the date of the cosmic ray event that caused the carbon-14 content anomaly in tree rings to be March 774 ± 6 months in 1 sigma range. Therefore, the date of "white rainbow" on June 775, which was recorded in the Japanese literature, is well inside the uncertainty range of Guettler's estimation.

Here we do not attempt to contest that the carbon-14 event in 775 was indeed of the solar origin. More evidence is definitely necessary to claim that the "white rainbow" in the Japanese literature was an aurora, and hence, we cannot draw any conclusive remark on the origin of the carbon-14 event. However, large-scale solar flares should cause simultaneous observation of auroras over a large area. Further investigation on the existence of auroral records on the nearby date would help in identifying the reason behind the rapid increase in carbon-14 in tree rings.

**Conclusion**

In this study, we examined the usages of "white rainbow（白虹）" and "unusual rainbow (虹蜺)" in the treatises of the Chinese official histories during the reign of *Táng* (618 to 907). These words were used to express heavenly events, and it was shown that their usages are consistent with the interpretation of auroras. However, this does not mean that all "white rainbow（白虹）" and "unusual rainbow (虹蜺)" events were auroras. The probability of being auroras is enhanced if there were simultaneous records of similar events at distant places. Although we have not found such records in the official histories of *Táng*, we found an example of simultaneous observations in China and Japan, where "white rainbow" was used in the Chinese record. Our results strongly suggest that a re-examination of oriental historical sources is required to make a more comprehensive survey of pre-modern auroras observations, not only in China but also in other areas under the cultural influence of China, such as Japan or Korea.

**Authors' contributions**

This research was performed with the cooperation of the following authors: HH made philological and historical contributions. HI, ADK, and HT made astronomical contributions. RK and HM helped with the interpretation and helped draft the manuscript. All authors have read and approved the final manuscript.

**Acknowledgements**

We thank anonymous referee for interesting comments and suggestions.

This work was supported by the Center for the Promotion of Integrated Sciences (CPIS) of SOKENDAI as well as the Kyoto University's Supporting Program for Interaction-based Initiative Team Studies "Integrated study on human in space" (PI: H. Isobe), the Interdisciplinary Research Idea contest 2014 by the Center of Promotion Interdisciplinary Education and Research, the "UCHUCHUGAKU" project of the Unit of Synergetic Studies for Space, the Exploratory Research Projects of the Research Institute of Sustainable Humanosphere, Kyoto University, and Grants-in-Aid

**Appendix: Hard copy information for historical sources**

Here we provide the publication information on the hard copies of the historical sources with full respect for the manner of historians.

*Gogumaiki*: Sanjo Kimitada, The Historiographical Institute of the University of Tokyo (ed.) (1980) *Gogumaiki* (後愚昧記), I-IV, Tokyo: Iwanami Shoten [published in Japanese].

*Jiùtángshū*: Liú Xù (1975) *Jiùtángshū* (舊唐書), I-XVI, Běijīng: Zhōnghuá Shūjú [published in Chinese].

*Nihonshoki*: Kuroita K (ed.) (1967) *Nihonshoki* (日本書紀), vol. I-III, Tokyo: Iwanami Shoten [published in Japanese].

*Shokunihongi*: Kuroita K (ed.) (1966) *Shokunihongi* (続日本紀), Tokyo: Yoshikawa Kobunkan [published in Japanese].

Tiānyuán Yùlì Xiángyìfù: "*Tiānyuán Yùlì Xiángyìfù* (天元玉曆祥異賦)", a manuscript at 305-257, Naikaku Bunko, Books of Shoheizaka Gakumonjo, in the National Archives of Japan (国立公文書館 昌平坂學問所本 內閣文庫 305-257) [written in Chinese]

*Xīntángshū*: Ōu Yángxiū, Sòng Qí (1975) *Xīntángshū* (新唐書), I-XX, Běijīng: Zhōnghuá Shūjú [published in Chinese].

*Yuánshǐ*: Sòng Lián (1976) *Yuánshǐ* (元史), I-XV, Běijīng: Zhōnghuá Shūjú [published in Chinese].

**References**

Allen, J., Lou, F., Herb, S. , & Patricia R 1989, *EOS Transactions, 70*, 1479

Araki, T. 2014, *Earth, Planets, and Space* 66:164

Butler, R. F. 1992, *Paleomagnetism: magnetic domains to geologic terranes*. Oxford: Blackwell Scientific Publications.

Carrington, R. C. (1871) A solar fog-bow. *Mon. Not. R. Astron. Soc*, 31, pp. 74-75.

Chapman, J., Neuhäuser, D., Neuhäuser, R., & Csikszentmihalyi M. 2015, *Astronomical Notes*, 336, 530.

Eddy, J. A. 1980, The historical record of solar activity. In: The ancient sun: Fossil record in the earth, moon and meteorites; Proceedings of the Conference, Boulder, CO, October 16-19, 1979. (A81-48801 24-91).( *Pergamon Press, New York and Oxford)* . 119

Fritz H (1873) *Verzeichniss Beobachteter Polarlichter.* C. Gerold's & Sohn, Wien. OpenURL

Guettler H et al. 2015, *Earth and Planetary Science Letters*, 411, 290

Hambaryan, V. V., & Neuhäuser, R. 2013, *MNRAS*, 430, 32

Hayakawa, H., Tamazawa, H., Kawamura, A. D., & Isobe, H. 2015  *Earth, Planets, and Space,* 67: 82

Kanda, S. 1933,  *Astron Her* 26 (11): 204-210

Keimatsu, M. 1970, *Annals of Science of Kanazawa University* 7: 1-10

Keimatsu, M. 1971, *Annals of Science of Kanazawa University* 8: 1-14

Keimatsu, M. 1972, *Annals of Science of Kanazawa University* 9: 1-36

Keimatsu, M. 1973, *Annals of Science of Kanazawa University* 10: 1-32

Keimatsu, M. 1974, *Annals of Science of Kanazawa University* 11: 1-36

Keimatsu, M. 1975, *Annals of Science of Kanazawa University* 12: 1-40

Keimatsu, M. 1976, *Annals of Science of Kanazawa University* 13: 1-32

Liu, Y. et al. 2014, *Nature Scientific Reports*, 4 (3728).

Maehara, H., Shibayama, T., Notsu, S., Notsu, Y., Nagao, T., Kusaba, S., Honda, S., Nogami, D., & Shibata, K. 2012, *Nature*, 485: 478-481

Mekhaldi, F. et al. (2015) Multiradionuclide evidence for the solar origin of the cosmic-ray events of AD 774/5 and 993/4, *Nature Communications* 6, Article number: 8611 doi:10.1038/ncomms9611

Miyake, F., Nagaya, K., Masuda, K., & Nakamura, T. 2012, *Nature*, 486: 240–242.

Miyake, F., Suzuki, A., Masuda, K., Horiuchi, K., Motoyama, H., Matsuzaki, H., Motizuki, Y., & Nakai, Y. 2015, *Geophysical Research Letters*, 42, 1: 84-89.

Nakazawa, Y., Okada, T., & Shiokawa, K. 2004, *Earth, Planets, and Space*, 56: e41-e44.

Neuhäuser, R., & Neuhäuser, D. L. 2015, *Astronomische Nachrichten*, 336, 3: 225.

Owens, B. 2013, *Nature*, 495: 300-303.

Pavlov, A. K., Blinov, A. V., Konstantinov, A. N., Ostryakov, V. M., Vasilyev, G. I., Vdovina, M. A., & Volkov, P.A. 2013, *MNRAS*, 435: 2878–2884.

Ramus, J. F. 1745, *Acta Societatis Hafniensis*, Copenhagen.

Shibata, K. et al. 2013, *Publ Astron Soc Jpn* 65: 49S.

Shibayama, T., Maehara, H., Notsu, S., Notsu, Y., Nagao, T., Honda, S., Ishii, T. T., Nogami, D., & Shibata, K. 2013, *ApJS*, 209 (5): 13.

Shiokawa, K., Ogawa, T., & Kamide, Y. 2004, *J Geophys Res* 110 (9):A05202.

Usoskin, I. G. 2013, *Living Rev Solar Phys*, 10: 1.

Vaquero, J.M., et al. (2002) An Observation of a Fog Bow in the Natural Park of Monfragüe, Spain. *Weather* 57 (12), 446-448.

Willis, D. M., & Stephenson, F. R. 1999, *Annales Geophysicae*, 18: 1-10

Yau, K. K. C., Stephenson, F. R., & Willis, D. M. 1995, *Council for the central laboratory of the research councils*. Technical Report RAL-TR-95-073.

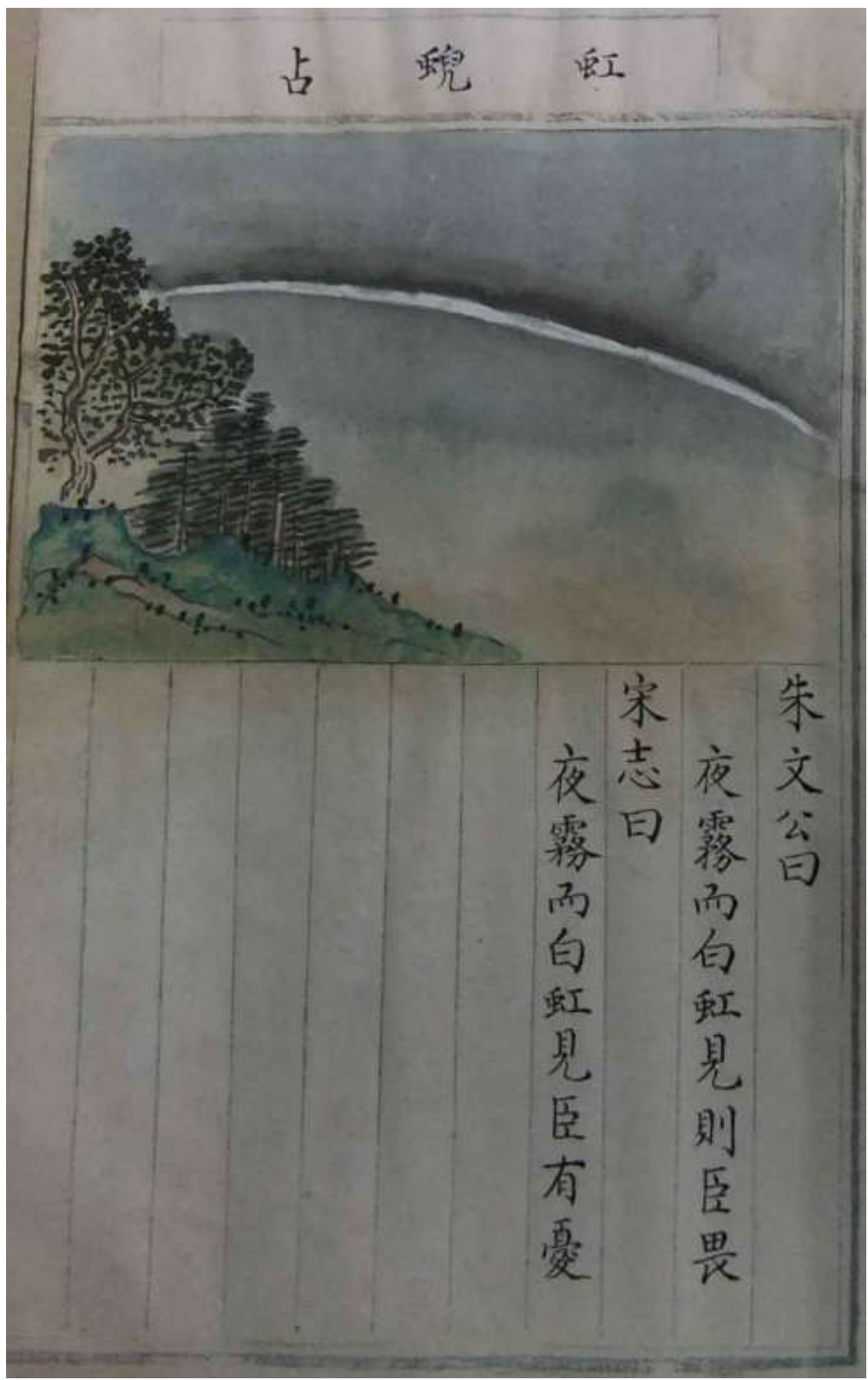

Fig1: A historrical figure of "white rainbow" in "Tiānyuán Yùlì Xiángyìfù (天元玉曆祥異賦)"

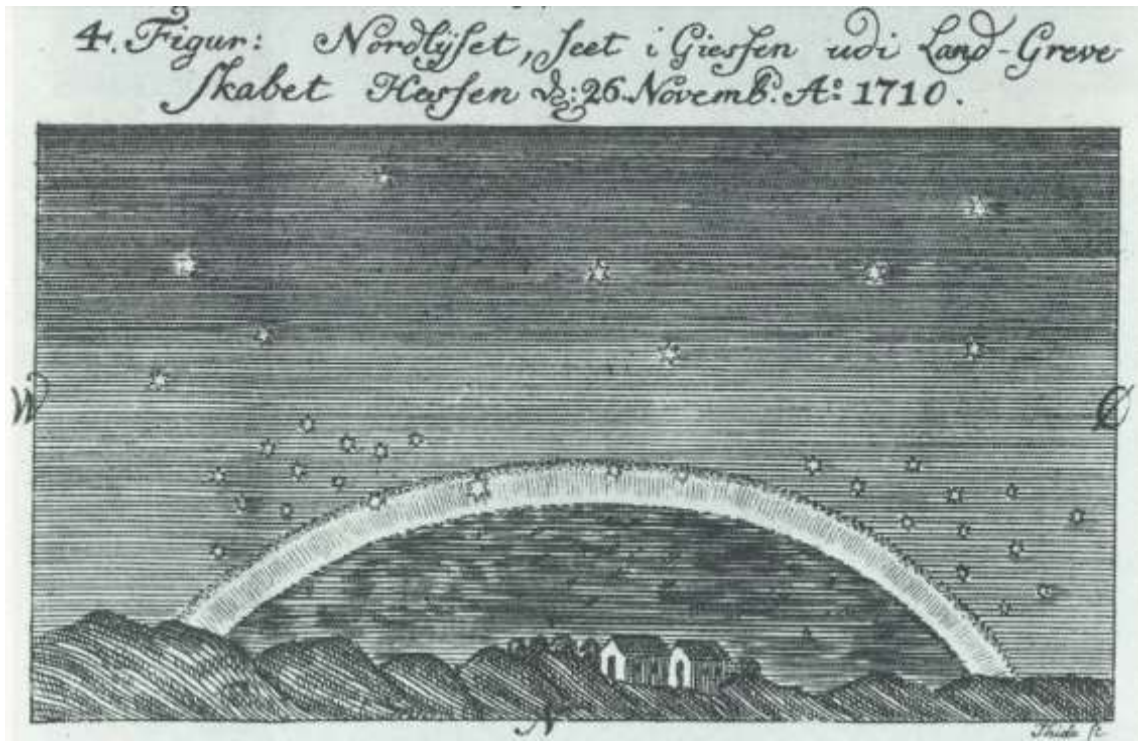

Fig.2 A historical figure of bow-shaped aurora in 26 Nov 1710 in Ramus (1745)